\documentclass[letter,11pt]{article}
 \usepackage{jheppub}
 
 \usepackage{graphicx}
 \usepackage{hyperref}
 \usepackage[utf8]{inputenc}
 \usepackage{amssymb}
\usepackage{multirow}
\usepackage{soul}
 
 \usepackage{booktabs}
\usepackage{mathrsfs}
\usepackage{epsfig}
\usepackage{graphicx}
\usepackage{subfigure}
\usepackage{dcolumn}
\usepackage{bm}
\usepackage{amsmath}
\usepackage{slashed}
\usepackage{multirow}
\usepackage{color}
\usepackage{dcolumn}
\usepackage{bm}
\usepackage{color}

\allowdisplaybreaks

 \newcommand{\lsim}{{\;\raise0.3ex\hbox{$<$\kern-0.75em\raise-1.1ex\hbox{$\sim$}}\;}}
\newcommand{\gsim}{{\;\raise0.3ex\hbox{$>$\kern-0.75em\raise-1.1ex\hbox{$\sim$}}\;}}
\newcommand{\beq}{\begin{equation}}
\newcommand{\eeq}{\end{equation}}
\newcommand{\bea}{\begin{eqnarray}}
\newcommand{\eea}{\end{eqnarray}}

\def\baa{\begin{array}}
\def\eaa{\end{array}}

\mathchardef\minus="002D

\preprint{ }

\title{Correlation length of the angular mode for an approximate $U(1)$ symmetry during inflation}

\author{Chengcheng Han}
\emailAdd{hanchch@mail.sysu.edu.cn }
\affiliation{School of Physics, Sun Yat-Sen University, Guangzhou 510275, China}

\vspace{0.5cm}

\abstract{
It is known that a light scalar field obtains fluctuations in the de Sitter inflationary background. Such fluctuations could provide an initial condition for baryogenesis through the Affleck-Dine mechanism, where an approximate $U(1)_B$ symmetry is usually assumed. However, an interpretation of the baryon number generation in this way is strongly related to the correlation length of the angular mode. In this work, we calculate the correlation length of the angular mode for a model exhibiting an approximate $U(1)$ symmetry. We find that for a massive nearly non-interacting field, the correlation length of the angular mode is determined by the mass parameter of the model and it is similar to $H^{-1} \exp(H^2/m^2)$. Applying this result to baryogenesis via the Affleck-Dine mechanism with a stochastic origin, we find that only for $m \ll \mathcal O(0.1) H$(assuming $N_*=60$) can the correlation length of the baryon number density be much larger than our current horizon size, such that we live in the baryon-rich region. If this is not true, at early times our universe would consist of numerous patches of  baryon-rich and anti-baryon-rich regions with the average baryon number being nearly zero. 
 }

\makeatletter
\def\@fpheader{\relax}
\makeatother

\begin{document} 
\maketitle
\flushbottom
\newpage

\section{Introduction}

It is widely believed the universe undergoes fast expansion in its early stage (inflation), characterized by a quasi-de Sitter space-time~\cite{Brout:1977ix, Sato:1980yn, Guth:1980zm, Linde:1981mu, Albrecht:1982mp}. The period of inflation not only solves the flatness problem, the horizon problem, but also seeds the primordial fluctuations of the Cosmic Microwave Background (CMB) temperature, which finally resulted in the large structure of our universe. During this stage, all light scalar fields obtain fluctuations with an amplitude around $\frac{H}{2\pi}$ in each e-fold, and the distribution function of the field values can be described by the Fokker-Planck equations. For sufficient long inflationary time, the fluctuations continue to accumulate, and finally, the distribution function reaches an equilibrium state~\cite{Bunch:1978yq, Starobinsky:1994bd}. For a massive nearly non-interacting field, the equilibrium solution is characterised by a correlation length of around $H^{-1} \exp(H^2/m^2)$ within which we can take the field value as a constant~\cite{Bunch:1978yq, Starobinsky:1994bd}. This equilibrium distribution provides an initial condition for the vacuum displacement of the scalar field. If this field carries a $U(1)_B$ or $U(1)_L$ charge, it may generate the baryon asymmetry of our universe through the Affleck-Dine mechanism~\cite{Affleck:1984fy, Dine:1995uk, Dine:1995kz}. However, due to the stochastic behavior of the field, the average baryon number in the whole universe vanishes. But it has been argued that if $m \ll H$~\cite{ Dine:1995kz, Linde:1985gh}, the correlation length of the field is much larger than our current horizon size, such that we could live in a baryon-rich patch.  

Recently, some studies have tried to use similar ideas to generate the baryon asymmetry via the Affleck-Dine mechanism~\cite{Wu:2020pej, Wu:2021mwy, Wu:2021gtd}. However, to avoid the baryonic isocurvature limit from CMB observation, they require the mass of the scalar field to be close to the inflationary Hubble parameter. Since the correlation length of the field is around $H^{-1} \exp(H^2/m^2)$, in the early time our current universe should consist of numerous Hubble patches with different baryon numbers. Nevertheless, the observed baryon number of our universe should be the average of all patches. Now we come up with a question: whether the average of the baryon number in our universe (close to) zero or not? 

This question is closely related to the correlation length of the angular mode $\theta$. In the Affleck-Dine mechanism, the baryon number asymmetry is usually proportional to $\sin n\theta$, where the value of $n$ depends on the models. Due to the stochastic behavior of the field, $\theta$ is randomly distributed and the average of the baryon number in the whole universe should be zero. However, if the correlation length of $\theta$ is much larger than our current horizon size, we can still live in a baryon-rich patch, even if the average of the baryon number in the whole universe is zero. On the other hand, if the correlation length of $\theta$ is much smaller than our current horizon size, then at early times our universe consists of many baryon-rich patches and anti-baryon-rich patches with the same probabilities, then the average of the baryon number in our universe is nearly zero\footnote{It can not be exactly zero. According to the central limit theorem, the average of the baryon number in our universe should be around $|Y_{B}| e^{-3/2 N}$ where $Y_B$ is the baryon to entropy ratio defined as $\sqrt{\langle Y_B^2\rangle } $ and $N \sim N_*/\log(HR_c)$ is the number of the patches contained in our current horizon. Here $N_*\approx 60$ is the e-fold number of our observed universe.}. 

Since the correlation length for a weakly coupled scalar field depends on its mass, one may argue that for a small $U(1)_B$ breaking term, the angular mode would have an effective ``mass" much smaller than the Hubble parameter. In this case, one may conclude that the correlation length of the $\theta$ could be much larger than our current horizon size. A similar argument has been used in the works~\cite{Wu:2020pej, Wu:2021mwy, Wu:2021gtd}\footnote{See the discussion of \cite{ Wu:2021mwy} at the end of Sec. 3.}.

\begin{figure}
      	\centering
      	\includegraphics[width=0.4\textwidth]{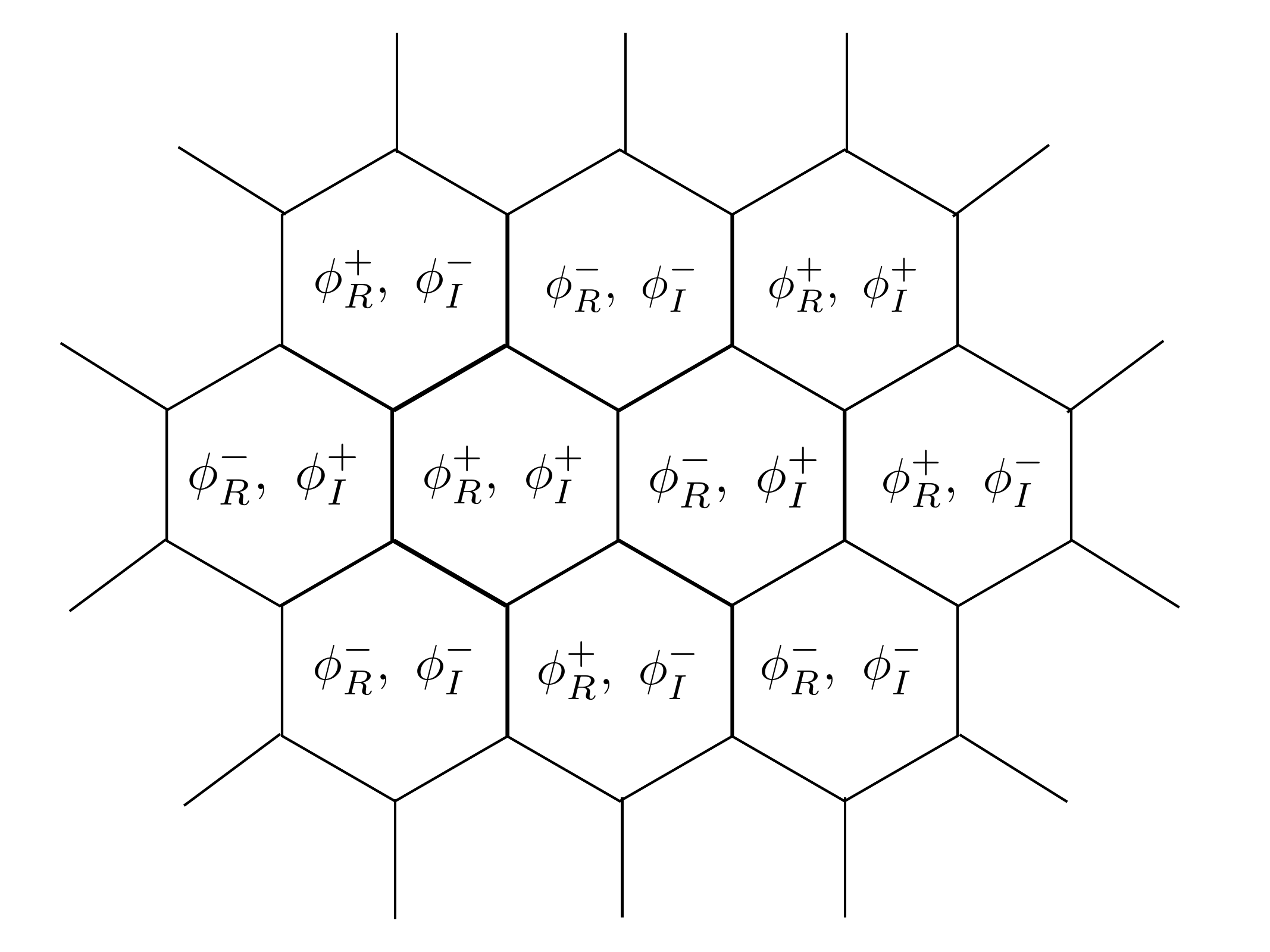}
      	\caption{Illustration of the  correlation length for the angular mode.  Within each cell, the  $\phi_R,\phi_I$ can be taken as a constant, with the size of each cell is around $R_c$. Here $\phi_{R,I}^+$($\phi_{R,I}^-$) denotes the region with $\phi_{R,I} > 0$($\phi_{R,I} < 0$) and $\tan \theta = \frac{\phi_I}{\phi_R}$. }
      	\label{cell}
      \end{figure}

On the other hand, since the Affleck-Dine field contains two real fields $\phi_R$ and $\phi_I$ with similar correlation length $R_c \approx H^{-1} \exp(H^2/m^2)$, one can imagine that the whole universe is divided by many cells with a scale of $R_c$. Within each cell, we can take $\phi_I$ and $\phi_R$ as a constant, but for different cells, the field values of $\phi_{R, I}$ are not strongly correlated and they can be taken as random distributed\footnote{The distribution functions are decided by the equilibrium solution of the Fokker-Planck equations.}. Due to the approximate $U(1)$ symmetry, the sign of $\phi_R$ and $\phi_I$ are also random with equal probabilities to be positive or negative. Note that $\tan \theta = \frac{\phi_I}{\phi_R}$, so the sign of $\tan \theta$ is also randomly distributed. Therefore, we expect that the correlation length of $\theta$ should not be too much larger than $R_c$, as illustrated in Fig.~\ref{cell}.

To resolve this contradiction, in this work we implicitly calculate the correlation length of $\theta$. We find that if the model presents an approximate $U(1)$ symmetry, the correlation length of the angular mode is indeed not much larger than $R_c$. The paper is organized as follows, in Sec.~\ref{sec1} we show the general formalism for the calculation, in Sec.~\ref{sec2} we take the single field as one typical example, and we calculate the correlation length of angular mode for the exact $U(1)$ case in Sec.~\ref{sec3} and approximate $U(1)$ case in Sec.~\ref{sec4}. We also calculate the correlation length of the baryon number density in Sec.~\ref{sec5} and we discuss the results in Sec.~\ref{sec6}.

\section{General formalism}
\label{sec1}
In this section, we present the general formalism for our calculation. This part is based on the work~\cite{Starobinsky:1994bd} and we just summarize the main result here. During inflation, the scalar fields follow a stochastic process, which can be described by the Fokker-Planck equation,
\begin{eqnarray}
\frac{\partial \rho(\varphi,t)}{\partial t} =\frac{1}{3H} \left[  \rho(\varphi,t) \nabla^2 V(\varphi) + \nabla V(\varphi) \cdot \nabla \rho(\varphi,t)   \right] +\frac{H^3}{8\pi^2} \nabla^2 \rho(\varphi,t)~,
\end{eqnarray}  
where $V(\varphi)$ is the potential for $\varphi$ and $\rho(\varphi,t)$ is the probability distribution function. Here $\nabla$ only acts on the field space. 

To solve the Fokker-Planck equation, one can assume,
\begin{eqnarray}
\rho(\varphi,t)= \exp\left(- \frac{4\pi^2 V(\varphi)}{3 H^4}\right) \sum_{n=0}^{\infty} a_n \Phi_n(\varphi) e^{-{\Lambda_n}(t-t_0)}~.
\end{eqnarray}  
Define
\begin{eqnarray}
v(\varphi)\equiv \frac{4\pi^2}{3 H^4} V(\varphi)~,
\end{eqnarray}  
then $\Phi_n$ satisfies the following eigenvalue equation,
\begin{eqnarray}
\frac{1}{2} \left[ - \nabla^2   +  \left ({\nabla v(\varphi)}^2- \nabla^2 v(\varphi) \right ) \right ] \Phi_n(\varphi)= \frac{4\pi^2}{H^3} \Lambda_n \Phi_n(\varphi)~.
\end{eqnarray}  
The coefficients $a_n$ are given by the initial condition of $\rho$ at $t=t_0$ as
\begin{eqnarray}
a_n= \int d\varphi \rho(\varphi,t_0) e^{v(\varphi)} \Phi_n(\varphi)~.
\end{eqnarray}  
Note that there is always a solution which is called the static equilibrium solution with $\Lambda_0=0$.
\begin{eqnarray}
\Phi_0(\varphi)= \mathcal N^{-1/2} e^{-v(\varphi)}~,
\end{eqnarray} 
where $\mathcal N$ is calculated by normalization condition
\begin{eqnarray}
\mathcal N \equiv  \int e^{-2 v(\varphi)} d\varphi~.
\end{eqnarray} 
After enough time has passed, any solution will finally approach the static equilibrium solution,
\begin{eqnarray}
\rho_{q}(\varphi) \equiv \mathcal N^{-1} \exp\left( -\frac{8\pi^2}{3H^4}V(\varphi) \right) =\mathcal N^{-1} e^{-2 v(\varphi)}~.
\end{eqnarray}  
We can calculate the average of any variable $F(\varphi)$ and $F^2(\varphi)$ in terms of the equilibrium solution,
\begin{eqnarray}
\langle F  \rangle = \int d\varphi F(\varphi) \rho_q(\varphi)~,  
\end{eqnarray} 
and
\begin{eqnarray}
\langle F^2  \rangle = \int d\varphi F^2(\varphi) \rho_q(\varphi)~.
\end{eqnarray} 

On the other hand, in the equilibrium state, we can derive the correlation function for any variable $F(\varphi)$,
\begin{eqnarray}
G_F(|t_1-t_2|) \equiv \langle  F(\varphi({\bf x}, t_1))  F(\varphi({\bf x}, t_2))  \rangle~.
\end{eqnarray}  
We have
\begin{eqnarray}
G_F(t) = \int F(\varphi)  \Xi(\varphi,t) d\varphi~,
\end{eqnarray}  
where 
\begin{eqnarray}
 \Xi(\varphi,t) \equiv \int F(\varphi_1) \rho_{q}(\varphi_1) \Pi[\varphi(t)|\varphi_1(0)] d\varphi_1~,
\end{eqnarray}  
satisfying the Fokker-Planck equation with the initial condition,
\begin{eqnarray}
 \Xi(\varphi,0) =F(\varphi) \rho_{q}(\varphi)~,
\end{eqnarray} 
and $\Pi[\varphi_2(t)|\varphi_1(0)]$ is the conditional probability to find $\varphi(t)=\varphi_2$ provided that $\varphi(0)=\varphi_1$.
Since $ \Xi(\varphi,t)$ satisfies the Fokker-Planck equation, it can be written as,
\begin{eqnarray}
\Xi(\varphi,t)= \exp\left(- \frac{4\pi^2 V(\varphi)}{3 H^4}\right) \sum_{n=0}^{\infty} A_n \Phi_n(\varphi) e^{-{\Lambda_n}t}~,
\end{eqnarray}  
where
\begin{eqnarray}
 A_n &=& \int F(\varphi) \rho_{q}(\varphi) e^{v} \Phi_n(\varphi) d \varphi = \mathcal N^{-1} \int F(\varphi) e^{-v} \Phi_n(\varphi) d \varphi \nonumber \\
 &=& \mathcal N^{-1/2} \int F(\varphi) \Phi_0 \Phi_n(\varphi) d \varphi~. 
\end{eqnarray}  
Then we have
\begin{eqnarray}
G_F(|t|) &=&\sum_{n=0} A_n e^{-\Lambda_n t} \int F(\varphi) e^{-v} \Phi_n(\varphi) d \varphi = \mathcal N \sum_{n=0} A_n^2 e^{-\Lambda_n t} 
= \sum_{n=0} \tilde A_n^2 e^{-\Lambda_n t}~,
\end{eqnarray}  
with
\begin{eqnarray}
 \tilde A_n &=& \int F(\varphi) \Phi_0(\varphi) \Phi_n(\varphi) d \varphi~.
\end{eqnarray}  
For a long enough period of time, the dominant contribution of $G_F(t)$ comes  from the lowest state $\Lambda_{\rm min}$ with $\tilde A_{\rm min} \ne 0$. We can use the above formula to calculate the average of the variable $F$ and $F^2$,
\begin{eqnarray}
 \langle  F \rangle &=&\left( \int F(\varphi) \rho_{q} d \varphi  \right) = \left(\int F(\varphi) \Phi_0(\varphi) \Phi_0(\varphi) d \varphi  \right)= \tilde A_0~, \nonumber \\
{\langle  F^2 \rangle} &=&G_F(0)=\sum_{n=0} \tilde A_n^2~.
\end{eqnarray} 
If we define the fluctuations of $F$ as $\delta F= F-\langle F \rangle$, then
\begin{eqnarray}
{\langle (\delta F)^2 \rangle} = \langle  F^2 \rangle - \langle  F \rangle^2 =    \sum_{n=1} \tilde A_n^2~.
\end{eqnarray} 

For the spatial correlation function $\xi_F({\bf x})= \langle  F(\varphi({\bf x_1}, t)) F(\varphi({\bf x_2},t)) \rangle $(${\bf x = x_1- x_2}$),  in the large ${\bf |x|}$ limit, we just need to take $t\rightarrow 2 H^{-1} \log(a H {\bf |x|})$,
\begin{eqnarray}
\xi_F({\bf x}) &=& \sum_{n=0} \tilde A_n^2 (aH{\bf |x|})^{-2\Lambda_n H^{-1} }~.
\end{eqnarray} 
For the case of $\tilde A_0 =0$, we can define the correlation length that the variable  $F(\varphi)$ satisfies,
\begin{eqnarray}
\xi_F(R_c) &=& G_F(0)/2~.
\end{eqnarray} 
Then within the length scale $R_c$ we can take $F$ approximately as a constant. Usually this scale is decided by the lowest eigenvalue $\Lambda_{\min}$ with $\tilde A_{\rm min} \ne 0$.

\section{Example for a single field}
\label{sec2}
In this section, we will use the above formulas to calculate the example of a single scalar field where only a mass term is included.  These results are already known and there is nothing new here. But this result is important to allow us to extend to the case of multi-dimensions, particularly into 2-dimensions which we are interested in.
Now the potential is  
\begin{eqnarray}
V(\varphi) = \frac{1}{2} m^2 \varphi^2,~~~v(\varphi)\equiv \frac{4\pi^2}{3 H^4} V(\varphi)~.
\end{eqnarray}  
The eigenvalue equation is,
\begin{eqnarray}
\frac{1}{2} \left[ - \frac{\partial^2}{\partial \varphi^2}+\left ({v'(\varphi)}^2- v''(\varphi) \right ) \right ] \psi_n(\varphi)= \frac{4\pi^2}{H^3} \Lambda_n \psi_n(\varphi)~.
\end{eqnarray}  
We can define an effective potential as,
\begin{eqnarray}
\mathcal V= \frac{1}{2} \left(   v'(\varphi)^2- v''(\varphi)  \right) = \frac{1}{2}\omega^2 \varphi^2-\frac{1}{2} \omega~,
\end{eqnarray}  
where $\omega= \frac{4\pi^2 m^2}{3 H^4} $. Then the eigenvalue equation becomes,
\begin{eqnarray}
\frac{1}{2} \left[ - \frac{\partial^2}{\partial \varphi^2}+\omega^2 \varphi^2 \right ] \psi_n(\varphi) -\frac{1}{2} \omega \psi_n(\varphi)= \frac{4\pi^2}{H^3} \Lambda_n \psi_n(\varphi)~.
\end{eqnarray}  
This is nothing but the eigenvalue equation for an oscillator with a frequency $\omega$. The difference is that the zero energy has been removed by the terms $-\frac{1}{2} \omega \psi_n(\varphi)$. One can easily find that the eigenvalues are,
\begin{eqnarray}
\Lambda_n =  \frac{H^3}{4\pi^2} n \omega =  \frac{n m^2}{3H},~ n=0,1,2...   ~.
\end{eqnarray}  
The eigenfunctions can be written as,
\begin{eqnarray}
\psi_n =  N_n \exp\left( -\frac{1}{2} \alpha^2 \varphi^2  \right) H_n(\alpha \varphi), ~~ N_n = (\alpha/\sqrt{\pi} 2^n n!)^{1/2}~,
\end{eqnarray}  
with $\alpha=\sqrt{\omega}$ and $H_n$ is the Hermite polynomials. For simplicity, we can use the Dirac Bra-ket notation to describe the eigenfunction  $\psi_n$ as  $|n\rangle$ with the eigenvalue  $\Lambda_n$. We also have the following relation,
\begin{eqnarray}
\varphi |n\rangle = \frac{1}{\alpha} \left[   \sqrt{\frac{n}{2}} |n-1 \rangle  +      \sqrt{\frac{n+1}{2}} |n+1 \rangle   \right]~.
\end{eqnarray}  
It is then not difficult to find that for  $F(\varphi) =\varphi^2$, we have
\begin{eqnarray}
\langle F \rangle =\langle \varphi^2 \rangle =\langle 0 |\varphi^2 |0\rangle = \frac{1}{2\alpha^2} = \frac{3H^4}{8\pi^2m^2}~. 
\end{eqnarray}  
For $F(\varphi) =\varphi$, we have
\begin{eqnarray}
\langle 0 |\varphi |0\rangle = 0,~~\langle 0 |\varphi |1\rangle = \frac{1}{\sqrt{2}\alpha}~ =\tilde A_1
\end{eqnarray}  
The spatial correlation function of $F(\varphi) = \varphi$ is
\begin{eqnarray}
\xi({\bf x}) = |\tilde A_1|^2 (aH{\bf |x|})^{-2\Lambda_1 H^{-1}} = \langle \varphi^2 \rangle (aH{\bf |x|})^{-2\Lambda_1 H^{-1}}~.
\end{eqnarray}  
Then we have
\begin{eqnarray}
R_c= H^{-1} \exp{\left( \log2/ 2\Lambda_1\right)}= H^{-1} \exp{\left(   \frac{3 {\rm \log} 2}{2} \frac{H^2}{m^2}   \right)} \approx H^{-1}\exp{\left(\frac{H^2}{m^2}   \right)}~,
\end{eqnarray}  
therefore, the correlation length is totally decided by the mass of the field(we set the $a=1$ at the end of inflation). For $N_*= 60$, if the mass is $m \lesssim 0.1 H$, the correlation length $\varphi$ can be much larger than our observed universe.  If $m \gtrsim 0.13 H$(assuming $N_*=60$), in the early time our current universe is composed of different patches with different $\varphi$ following a distribution described by $\rho_q(\varphi)$.

\section{Correlation length of the angular mode for an exact $U(1)$ symmetry}
\label{sec3}
In this section, we will calculate the correlation length of the angular mode for an exact $U(1)$ symmetry. In this case, the $\theta$ is massless and the correlation length may appear to be very large. We will show that this naive estimation is wrong. For multi-dimension Fokker-Planck equation,
\begin{eqnarray}
-\frac{1}{2} \nabla^2 \Phi_{n} +\frac{1}{2} \left[ (\nabla v)^2 -\nabla^2 v \right] \Phi_{n}=\frac{4\pi^2\Lambda_n}{H^3} \Phi_{n}~.
\end{eqnarray}
Let us look at a simple case,
\begin{eqnarray}
V(\phi) &=& m^2 |\phi|^2 = \frac{1}{2}m^2 \phi_1^2 + \frac{1}{2}m^2 \phi_2^2~,
\end{eqnarray}
where $\phi = \frac{1}{\sqrt{2}}(\phi_1 + i\phi_2) $. From our previous calculation, it shows that the universe is divided into different patches with a correlation length of $R_c \approx H^{-1}\exp \left(\frac{H^2}{m^2}  \right)$. However,  if we use the radial-angular coordinates $\phi= \frac{r}{\sqrt{2}} e^{i\theta}$, we have
\begin{eqnarray}
V(\phi) &=& m^2 |\phi|^2 = \frac{1}{2}m^2 r^2~.
\end{eqnarray}
Then we have one massive mode and one massless mode. One may argue that the correlation length for $\theta$ should be much larger than $R_c$. We will show that it is not true through two different coordinate representations.

\subsection{$\phi_1$, $\phi_2$  representation}
Since the $\phi_1, \phi_2$ do not couple to each other, the eigenvalues are easy to solve. The total eigenfunctions are just the product of each separate eigenfunction. Let us define the eigenfunction with eigenvalue  $\Lambda_{n, l}$ as
\begin{eqnarray}
\Psi_{n,l}(\phi_1,\phi_2)&=&\psi_n(\phi_1) \tilde \psi_l(\phi_2), ~\Lambda_{n, l}= \frac{(n+l)m^2}{3H},~ n,l=0,1,2...~.
\end{eqnarray}
There is a degeneracy for the eigenvalue $\Lambda_{n, l}$ with $N_{\Lambda_{n, l}} = n+l+1$. Here we only list some of the eigenfunctions with the lowest eigenvalues,
\begin{eqnarray}
\Psi_{0,0}(\phi_1,\phi_2)&=&\psi_0(\phi_1) \tilde \psi_0(\phi_2), ~\Lambda_0=0 \\
\Psi_{0,1}(\phi_1,\phi_2)&=&\psi_0(\phi_1) \tilde \psi_1(\phi_2), ~\Lambda_{0,1}=\frac{m^2}{3H} \\
\Psi_{1,0}(\phi_1,\phi_2)&=&\psi_1(\phi_1) \tilde \psi_0(\phi_2), ~\Lambda_{1,0}=\frac{m^2}{3H}~.
\end{eqnarray}
We can also use the Dirac Bra-ket notation $|n,l\rangle$ to denote the state with eigenfunction $\psi_n(\phi_1) \tilde \psi_l(\phi_2)$.

Let us calculate the correlation length of the angular mode $F(\phi_1,\phi_2)=\theta ={\rm Arcsin} \frac{\phi_2}{\sqrt{\phi_1^2+\phi_2^2}}$.  Due to the symmetry of the function, the first non-vanishing $\tilde A$ appears at 
\begin{eqnarray}
 \tilde A_{0,1} &=& \int  {\rm Arcsin} \frac{\phi_2}{\sqrt{\phi_1^2+\phi_2^2}} \Psi_{0,0}(\phi_1,\phi_2) \Psi_{0,1}(\phi_1,\phi_2) d \phi_1 d\phi_2~,
\end{eqnarray}  
with the corresponding eigenvalue $\frac{m^2}{3H}$. Numerically we find $A_{0,1} =0.8 \ne 0$,  then  $R_\theta \approx R_c$.

However, it is more intriguing to calculate the correlation length of $\sin \theta =\frac{\phi_2}{\sqrt{\phi_1^2+\phi_2^2}}$. Within the correlation length $R_\theta$, we can take $\theta$ as a constant, then the same is true for $\sin \theta$. Therefore, we expect the correlation length of these two variables might be similar (generally $f(\theta)$ and  $\theta$ could have different correlation lengths). More importantly, it is much easier to apply this variable to the $r-\theta$ representation. As usual,
\begin{eqnarray}
 \tilde A_{0,1} &=& \int \frac{\phi_2}{\sqrt{\phi_1^2+\phi_2^2}} \Psi_{0,0}(\phi_1,\phi_2) \Psi_{0,1}(\phi_1,\phi_2) d \phi_1 d\phi_2~.
\end{eqnarray}  
Numerically,  we find $\tilde A_{0,1} \approx 0.63$, and thus, the spatial correlation function is 
\begin{eqnarray}
\xi_{\sin \theta}({\bf x}) &=& \tilde A_1^2 (aH{\bf |x|})^{-2\Lambda_{0, 1} H^{-1}}~,
\end{eqnarray} 
where $\Lambda_{0, 1}= \frac{m^2}{3 H}$. Then we have $R_{\sin \theta }  = H^{-1}\exp \left(\frac{H^2}{m^2}  \right)$. Indeed the correlation length is same as $R_c$.

\subsection{$r$, $\theta$ representation}

For the  eigenvalue equation,
\begin{eqnarray}
-\frac{1}{2} \nabla^2 \Phi_{n} +\frac{1}{2} \left[ (\nabla v)^2 -\nabla^2 v \right] \Phi_{n}=\frac{4\pi^2\Lambda_n}{H^3} \Phi_{n}~,
\end{eqnarray}
we have 
\begin{eqnarray}
\nabla^2 =\frac{\partial^2 }{\partial r^2} +\frac{1}{r}\frac{\partial}{\partial r} +\frac{1}{r^2} \frac{\partial^2}{\partial \theta^2}~.
\end{eqnarray}
 in the $r-\theta$ representation. The effective potential $\mathcal V$ is only a function of $r$.
\begin{eqnarray}
\mathcal V= \frac{1}{2} \left[ (\nabla v)^2 -\nabla^2 v \right]~.
\end{eqnarray}  
Due to the rotation symmetry, the wave function can be written as,
\begin{eqnarray}
\Psi_{n,l}(r,\theta) = R_{n,l}(r) Y_l(\theta),~ l=0, \pm 1, \pm 2...~,
\end{eqnarray}
where 
\begin{eqnarray}
Y_l(\theta)= \frac{1}{\sqrt{2\pi}}e^{il\theta},~ l=0, \pm 1, \pm 2...~,
\end{eqnarray}
and we have 
\begin{eqnarray}
\int^{2\pi}_0 d\theta Y_l(\theta) Y^*_{l^\prime}(\theta)= \delta_{l,l^\prime}~.
\end{eqnarray}
Since $v$ is only a function of $r$, the eigenvalue equation becomes
\begin{eqnarray}
- \left[ R^{\prime \prime}_{n,l} + \frac{R^{\prime}_{n,l}}{r} - \frac{l^2}{r^2} R_{n,l} \right]+ 
\left[  (v^\prime)^2 - v^{\prime \prime}-\frac{v^\prime}{r} \right] R_{n,l}=\frac{8\pi^2\lambda_{n,l}}{H^3} R_{n,l}~,
\end{eqnarray}
where we have used the following relation 
\begin{eqnarray}
\frac{\partial^2}{\partial \theta^2}Y_{l}&=& -l^2~.
\end{eqnarray}
The eigensystem is similar to the 2 dimension oscillator
\begin{eqnarray}
 \lambda_{n,l} &=& (2n+|l|)\frac{m^2}{3H},~n=1,2,3...,~l=0, \pm 1,\pm 2...               \nonumber \\
  N_{\lambda_{n,l} } &=& 2n + |l|+1  ~.
\end{eqnarray}
This result is totally consistent with that found for the $\phi_1$,$\phi_2$ coordinates.
Here we list some of the wave-functions
\begin{eqnarray}
\Phi_{0,0}(r,\theta) &=& R_{0,0}(r) \frac{1}{\sqrt{2\pi}}   \nonumber \\
\Phi_{0,1}(r,\theta) &=& R_{0,1}(r)\frac{1}{\sqrt{2\pi}}e^{i\theta}  \nonumber \\
\Phi_{0,-1}(r,\theta) &=& R_{0,1}(r)\frac{1}{\sqrt{2\pi}}e^{-i\theta}~.
\end{eqnarray}
Actually it is more convenient to use another set of eigenfunctions,
\begin{eqnarray}
\Phi_{n,0}(r,\theta) &=& R_{n,0}(r) \frac{1}{\sqrt{2\pi}},~ n=0, 1,2,3...,  \nonumber \\
\Phi_{n,l}(r,\theta) &=& R_{n,l}(r)\frac{1}{\sqrt{\pi}} \cos l \theta ,~ n=0, 1,2,3...,~l=1,2,3... \nonumber \\
\Phi^\prime_{n,l}(r,\theta) &=& R_{n,l}(r)\frac{1}{\sqrt{\pi}}\sin l \theta,~ n=0, 1,2,3...,~l=1,2,3...~.
\end{eqnarray}
The lowest eigenvalue which has non-vanishing $  \tilde  {\mathcal A}$ is $\lambda_{0,1}= \frac{m^2}{3H} $ with 
\begin{eqnarray}
\tilde {\mathcal A}_{0,1}^\prime &=& \int \sin \theta \Phi_{0,0}(r,\theta) \Phi^\prime_{0,1}(r,\theta) rd r d\theta  \nonumber \\
 &=& \int R_{0,0}(r) R_{0,1}(r) rd r \frac{1}{\sqrt{2}}~.
\end{eqnarray}  
Numerically, we find  $\tilde {\mathcal A}^\prime_{0,1}\approx 0.63$, which is totally consistent with the previous calculation in which we used the $\phi_1, \phi_2$ coordinates.

\section{Correlation length of the angular mode for an approximate $U(1)$ symmetry}
\label{sec4}
For the case of an approximate $U(1)$ symmetry, we add a small breaking term for the $U(1)$ symmetry, which is a necessary condition for the baryogenesis by the Affleck-Dine mechanism. In this scenario, one may argue that for the approximate $U(1)$ symmetry, $\theta$ is not exactly massless and obtains a finite (small) mass, such that the correlation length could be much longer. In the following, we will show, that even if $U(1)$ is an approximate symmetry, the correlation length of $\theta$ is still close to $R_c$. 
Let us consider an example,
\begin{eqnarray}
V(\phi) &=& m^2 |\phi|^2 +\left[ \frac{1}{2}|\mu|^2\phi^2+ h.c  \right] \nonumber \\
&=& \frac{1}{2}(m^2+|\mu|^2) \phi_1^2 + \frac{1}{2}(m^2-|\mu|^2) \phi_2^2 \nonumber \\
&\equiv& \frac{1}{2}m_1^2 \phi_1^2 + \frac{1}{2}m_2^2 \phi_2^2 ~,
\end{eqnarray}
where the $|\mu|^2$ term breaks the $U(1)$ symmetry and we assume that such a term is much smaller than the $m^2$ term. In this case, we still have an exact solution.
\begin{eqnarray}
\Psi_{n,l}(\phi_1,\phi_2)&=&\psi_n(\phi_1) \tilde \psi_l(\phi_2), ~\Lambda_{n, l}= \frac{(n m_1^2+ l m_2^2)}{3H},~ n,l=0,1,2...~.
\end{eqnarray}
Here we list some of the wave functions with lowest eigenvalues, 
\begin{eqnarray}
\Psi_{0,0}(\phi_1,\phi_2)&=&\psi_0(\phi_1) \tilde \psi_0(\phi_2), ~\Lambda_0=0 \\
\Psi_{0, 1}(\phi_1,\phi_2)&=&\psi_0(\phi_1) \tilde \psi_1(\phi_2), ~\Lambda_1=\frac{m^2_2}{3H^2} \\
\Psi_{1,0}(\phi_1,\phi_2)&=&\psi_1(\phi_1) \tilde \psi_0(\phi_2), ~\Lambda_2=\frac{m^2_1}{3H^2} ~.
\end{eqnarray}
The degeneracy of the spectrum has been broken because $m_1 > m_2$. Now again, the lowest eigenvalue with non-vanishing $\tilde A$ is $\frac{m^2_2}{3H}$ with  
\begin{eqnarray}
 \tilde A_{0,1} &=& \int \frac{\phi_2}{\sqrt{\phi_1^2+\phi_2^2}} \Psi_{0,0}(\phi_1,\phi_2) \Psi_{0,1}(\phi_1,\phi_2) d \phi_1 d\phi_2~.
\end{eqnarray} 
For $|\mu|^2 \ll m^2$, the value of $\tilde A_{0,1}$ is essentially not changed. But the eigenvalue has been changed from $\frac{m^2}{3H^2}$ into $\frac{m^2_2}{3H^2}$ and the correlation length indeed becomes larger 
\begin{eqnarray}
R_{\sin \theta} \approx H^{-1}\exp \left({H^2}/{(m^2-|\mu|^2)} \right)~.
\end{eqnarray} 
Since we are considering the case of $\mu^2 \ll m^2$, this value is still close to $R_c$. More importantly, in the limit of $\mu^2 \rightarrow 0$, we get the exact $U(1)$ symmetry result.

To be complete, we also calculate the result in the $r,\theta$ coordinates. In this case we can use the formulas used in perturbation theory for quantum mechanics. We have
\begin{eqnarray}
V(\phi) &=& m^2 |\phi|^2 +\left[ \frac{1}{2}|\mu|^2\phi^2+ h.c  \right]  \nonumber \\
&=& \frac{1}{2} m^2 r^2 + \frac{1}{2}|\mu|^2 r^2 \cos 2 \theta~.
\end{eqnarray}
Since
\begin{eqnarray}
\Phi_0 &=& \mathcal N^{-1/2}  e^{-v}~,
\end{eqnarray}
obviously    
\begin{eqnarray}
\tilde A_0 = \int \Phi_0^2 \sin \theta r dr d\theta =0~.
\end{eqnarray}
 Note that $\Phi_0 \ne \Phi^{(0)}_{0,0}$ since $\Phi_0$ is the exact eigenfunction with the breaking term included. Now we need to find out the lowest eigenvalue with $\tilde A \ne 0$.
 The effective potential can be written as
\begin{eqnarray}
\mathcal V&=& \frac{1}{2} \left[ (\nabla v)^2 -\nabla^2 v \right]= \left[ \frac{1}{2} \left( \frac{4\pi^2}{3H^4} \right)^2  m^4 r^2   \frac{4\pi^2 m^2}{3H^4}  \right] + \left( \frac{4\pi^2}{3H^4} \right)^2  |\mu|^2 m^2 r^2 \cos 2\theta +\mathcal O(\mu^4)  \nonumber \\
&=& \left( {1}/{2} \omega_0^2 r^2-\omega_0 \right)+  \left(    \omega_0 \omega_1 r^2  \right) \equiv \mathcal V_0+ \mathcal V_b~,
\end{eqnarray}  
where we define $\omega_0=\frac{4\pi^2 m^2}{3H^4}$ and $\omega_1=\frac{4\pi^2 |\mu|^2}{3H^4}$. 
Clearly $\mathcal V_b$ breaks the degeneracy of the eigenvalues for states with   
\begin{eqnarray}
\Phi^{(0)}_{0,1}(r,\theta) &=& R^{(0)}_{0,1}(r)\frac{1}{\sqrt{\pi}} \cos \theta , \nonumber \\
{\Phi^\prime}^{(0)}_{0,1}(r,\theta) &=& R^{(0)}_{0,1}(r)\frac{1}{\sqrt{\pi}}\sin \theta~.
\end{eqnarray}
However, it does not mix these two states, and the eigenvalues of each state become,
\begin{eqnarray}
\lambda^b_{0,1} &=& \lambda_{0,1}^{(0)} + \frac{H^3}{4\pi^2} \int {{\Phi_{0,1}^{(0)}}}^2 \mathcal V_b r dr d\theta  \nonumber \\
{\lambda^\prime}^b_{0,1} &=& \lambda_{0,1}^{(0)} + \frac{H^3}{4\pi^2}\int {{{{\Phi^\prime}^{(0)}_{0,1}}}}^2 \mathcal V_b r dr d\theta~.
\end{eqnarray}
We have
\begin{eqnarray}
\frac{H^3}{4\pi^2} \int {{\Phi_{0,1}^{(0)}}}^2 \mathcal V_b r dr d\theta &=& \frac{\omega_0}{3H}   \frac{|\mu|^2}{2} \int {R^{(0)}_{0,1}}^2 {r^3} dr   \nonumber \\
\frac{H^3}{4\pi^2} \int {{{{\Phi^\prime}^{(0)}_{0,1}}}}^2 \mathcal V_b r dr d\theta &=&  -\frac{\omega_0}{3H} \frac{|\mu|^2}{2} \int {R^{(0)}_{0,1}}^2{r^3}  dr~. 
\end{eqnarray}
The last integration seems to be not easily determined. Notice that it is the average of $r^2= \phi_1^2 +\phi_2^2$ for the state with $\Lambda =\frac{m^2}{3H}$. We can use the $\phi_1$ $\phi_2$ coordinates to estimate this integration,  since we have
\begin{eqnarray}
\langle \Psi^{(0)}_{0,1} |\phi_1^2 +\phi_2^2 | \Psi^{(0)}_{0,1}  \rangle &=& \langle \Psi^{(0)}_{1,0} |\phi_1^2 +\phi_2^2 | \Psi^{(0)}_{1,0}  \rangle =\frac{2}{\omega_0}~.
\end{eqnarray} 
then we get 
\begin{eqnarray}
\int {R^{(0)}_{0,1}}^2 {r^3}  dr =  \frac{2}{\omega_0}~.
\end{eqnarray} 
In the end we find
\begin{eqnarray}
\lambda^b_{0,1} &=& (m^2 + |\mu|^2)  \frac{1}{3H} \nonumber \\
{\lambda^\prime}^b_{0,1} &=& (m^2 -|\mu|^2) \frac{1}{3H}~.
\end{eqnarray}Since we have $\mu^2 \ll m^2$, we do not expect $\tilde {\mathcal A}_{0,1}$ to change too much. The essential thing here is that the $\lambda$ has matched the value of the exact solutions in the  $\phi_1, \phi_2$ representation and the correlation length is determined by the eigenvalue. Since the correlation length is mainly determined by the lowest eigenvalue with $\tilde {\mathcal A} \ne  0$, we arrive at the same conclusion, that the correlation length of the angular mode should not be much larger than  $R_c$.

One can also extend this result to other small $U(1)$ breaking cases, from which we can then conclude that even if a small breaking term is included, the correlation length $\theta$ can not be much larger than $R_c$.

\section{Correlation length for the baryon number density}
\label{sec5}
It is known that a scalar field which carries baryon charge can generate a baryon asymmetry through the Affleck-Dine mechanism. In the following, we analyse the correlation length of the baryon number density.  The displacement of the field originates from quantum fluctuations. For the potential, we have
\begin{eqnarray}
V(\phi) &=& m^2 |\phi|^2 + \left[  \kappa_n \phi^n+ h.c  \right] ~,
\end{eqnarray}
and 
\begin{eqnarray}
\dot n_B + 3 H n_B = {\rm Im} (\phi \frac{\partial V}{\partial \phi})~.
\end{eqnarray}
The baryon number density at $H=m$ can be estimated as,
\begin{eqnarray}
n_{B} \sim \frac{n \kappa_n {\rm Im} (\phi^n) }{m}= \frac{n \kappa_n r^n \sin n\theta_0}{2^{n/2} m}~.
\end{eqnarray}
Since $\sin n\theta_0$ is randomly distributed, the average is given by $\langle  n_B  \rangle = 0$. But this just means that the whole universe has an average baryon number of zero, but not necessarily in our observed universe.  If the correlation length of $n_B$ is much larger than our current universe, we may still live in a region with a positive baryon number.  To calculate the correlation length, we set $n=4$ as an example,
\begin{eqnarray}
V(\phi) &=& m^2 |\phi|^2 +\left[ \kappa \phi^4+ h.c \right] \nonumber  \\
&=& \frac{1}{2} m^2 r^2 + \frac{1}{2}\kappa r^4 \cos 4 \theta ~.
\end{eqnarray}
The baryon number density is
\begin{eqnarray}
n_{B} = \frac{\kappa r^4 \sin 4\theta}{2 m} = \kappa \frac{4\phi_1\phi_2(\phi_1^2-\phi_2^2)}{2 m}~.
\end{eqnarray}
The correlation function is
\begin{eqnarray}
\langle n_{B}(0) n_B(t) \rangle = \sum \tilde A_n^2    e^{-\Lambda_n t}~,
\end{eqnarray}
where
\begin{eqnarray}
 \tilde A_n &=& \int n_B \Phi_0(\varphi) \Phi_n(\varphi) d \varphi~.
\end{eqnarray} 
Due to the $\sin 4\theta$ term, the major contribution is from   $\Phi_{0,4}$ with eigenvalue $\lambda_{0,4}= \frac{4m^2}{3H}$. Numerically, we find that
\begin{eqnarray}
 \tilde A_{0,4} &=& \int \frac{\kappa r^4 }{2 m} R_{0,0}(r)  R_{0,4} r dr \frac{1}{\sqrt{2}} =  \frac{1.8 \kappa }{m \omega^2}~.
\end{eqnarray}  
Then
\begin{eqnarray}
\langle n_{B}(0) n_B(t) \rangle \approx \frac{3\kappa^2}{m^2\omega^4} e^{-\frac{4m^2}{3H}t}~.
\end{eqnarray}
It is also useful to consider the $\phi_1, \phi_2$ coordinates.  For  $\Lambda= \frac{4m^2}{3H}$, there are five states,  
$| 0, 4 \rangle$, $| 4, 0 \rangle$, $| 1, 3 \rangle$,  $| 3, 1 \rangle$, $| 2, 2 \rangle$.   By defining $f(\phi_1, \phi_2)= 4\phi_1\phi_2(\phi_1^2-\phi_2^2)$, it is not difficult to find that
\begin{eqnarray}
\langle 0, 0| f(\phi_1, \phi_2) |4, 0 \rangle =\langle 0, 0| f(\phi_1, \phi_2) |0, 4 \rangle = \langle 0, 0| f(\phi_1, \phi_2) |2, 2 \rangle= 0~. 
\end{eqnarray}  
Then we only need to calculate,
\begin{eqnarray}
 |\langle 0, 0| f(\phi_1, \phi_2) |1, 3 \rangle| =  |\langle 0, 0| f(\phi_1, \phi_2) |3, 1 \rangle|  = \frac{\sqrt{6}}{\omega^2} ~.
\end{eqnarray}  
Now we have two non-vanishing $\tilde A$, which when summed together gives,
\begin{eqnarray}
\langle n_{B}(0) n_B(t) \rangle = \frac{3\kappa^2}{m^2\omega^4} e^{-\frac{4m^2}{3H}t}~.
\end{eqnarray}  
The spatial correlation of $n_B$ is
\begin{eqnarray}
\langle n_{B}(0) n_B({\bf x}) \rangle \approx \frac{3\kappa^2}{m^2\omega^4}(aH{\bf |x|})^{-\frac{8m^2}{3H^2}}~,
\end{eqnarray}  
which is consistent with the work~\cite{Hook:2015foa}.
Now we find that the correlation length of  $n_B$ is 
\begin{eqnarray}
R_{n_B} \approx H^{-1} \exp{ \left(   \frac{H^2}{4 m^2}  \right) }~.
\end{eqnarray}  
which is much smaller than $R_c$.  Note that even if $\langle n_B\rangle =0$, once  $R_{n_B} $ is much larger than our current universe, we could live in a baryon-rich universe, which is the case for  $m \ll 0.06 H$ for $N_*=60$.  

To end, we give one example of successful baryogenesis models despite the angular mode having a $R_c$ that  is much smaller than our current universe.  One such example is baryogenesis from sneutrino decay~\cite{Hamaguchi:2001gw}, the total baryon number is 
\begin{eqnarray}
{n_B}  = \epsilon n_\phi \approx {\epsilon} m |\phi|^2~.
\end{eqnarray}  
Since $n_B$ is independent of the $\theta$,  in the whole universe we could have the same sign of $n_B$. 

\section{Discussion and conclusion}
\label{sec6}
In this work, we calculated the correlation length of the angular mode for models with an approximate $U(1)$ symmetry. We find that for a massive nearly non-interacting fields, the correlation length of the angular mode is determined by the mass parameter, which is approximately $R_c=H^{-1} \exp(H^2/m^2)$. Applying this result to baryogenesis through the Affleck-Dine mechanism, if $m \sim H$, the correlation length of the angular mode would be much smaller than our current horizon size and the average of the baryon number in our universe would be nearly zero.

      \begin{figure}
      	\centering
      	\includegraphics[width=0.7\textwidth]{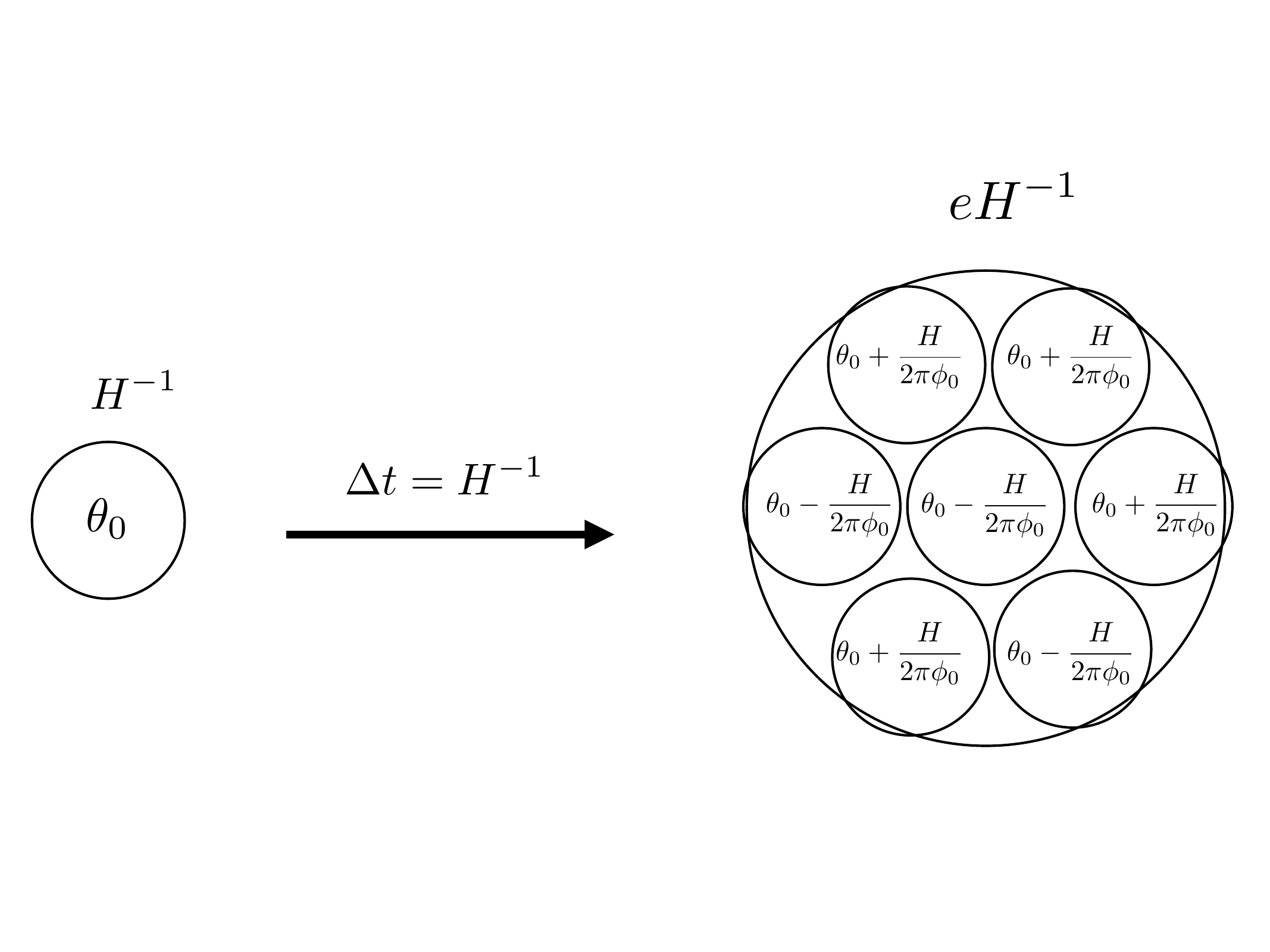}
      	\vspace{-1.5 cm}
      	\caption{Illustration of the  quantum jump for $\theta$.}
      	\label{theta}
      \end{figure}

 The reason for this is simple. 
 One can consider the quantum jump in each e-folding $\delta \phi \approx \frac{H}{2\pi}$ within a horizon size $H^{-1}$, then $\delta \theta \sim \frac{H}{2\pi \phi_0}$ where $\phi_0 \equiv \sqrt{\langle \varphi^2\rangle } =\sqrt{\frac{3}{2}} \frac{H^2}{2\pi m}$. 
 The breaking of the correlation length for $\theta_0\sim 1$ happens at an e-fold $N$ satisfying $\sqrt{N} = \frac{2\pi\phi_0}{H}\sim H/m$, therefore a natural correlation length for the $\theta_0$ is around $H^{-1} \exp \left( N\right) =H^{-1} \exp\left( \frac{H^2}{m^2}  \right)$ which is consistent with our calculation. For  $m \lesssim H$, we have $\phi_0 \sim H/2\pi$, then $\theta$ would soon become randomly distributed with a correlation length $H^{-1}$. Now we conclude that only when $m \ll \mathcal O(0.1) H$ (assuming $N_*=60$) can the correlation length of the baryon number density be much larger than our current horizon size, such that we live in the baryon-rich region\footnote{The baryonic isocurvature from CMB sets a stronger limit on $m$.}. 
 
In the work~\cite{Wu:2020pej}, the authors consider the stochastic origin of the initial condition for the Affleck-Dine mechanism. To avoid the baryonic isocurvature limit $\delta \theta /\theta \lesssim 10^{-5}$, usually one can take ${H}/{2\pi\phi_0} \lesssim 10^{-5}$ and ${m}/{H} \lesssim 10^{-5}$. However, the authors of~\cite{Wu:2020pej} argue that if $m\sim H$, the power spectrum of the baryon number is suppressed by an additional factor $\left( {k/}{H_{\rm end} } \right)^{-M_*}$ where $M_* = 2 m^2/3 H_*^2$. For $m \sim 0.5 H_*$, such suppression factor is large enough at the CMB length scale and the isocurvature limit can be avoided. But in this case the correlation length of angular mode is small and our universe will consist of many patches with positive baryon numbers and negative baryon numbers and the average of the baryon number in our observed universe is vanishing. It seems the authors do not realize this problem in their first paper~\cite{Wu:2020pej}. Later on, the authors combine the idea of ultra-slow-roll inflation(to generate the primordial black hole dark matter) with Affleck-Dine mechanism~\cite{Wu:2021gtd, Wu:2021mwy}. The ultra-slow-roll inflation provides an initial condition for the Affleck-Dine mechanism. Again, they choose a benchmark model of $m=0.5 H$ to generate the baryon asymmetry~\cite{Wu:2021gtd} to avoid the baryonic isocurvature bound. Until in their paper ~\cite{Wu:2021mwy}, they realize the importance of the correlation length $\theta$, then they add a paragraph to argue that the correlation length of $\theta$ is long(see the last paragraph of Sec. 3.2 in ~\cite{Wu:2021mwy}). The main argument is that $m^2_\theta \sim \partial_\theta \partial_\theta V(R,\theta) \rightarrow 0$ and the correlation length of $\theta$ is long due to the formula $R_c \sim H^{-1} \exp(H^2/m^2)$. However, since $\theta$ is not a canonical field and as we calculated in the work, the correlation length of $\theta$ is not decided by the $\theta$ mass, therefore their argument is wrong and their result is doubtable.

To end, we give a comment on the correlation length of $\theta$ in the original Affleck-Dine baryogenesis scenario, where the Affleck-Dine field gets a (large) 
negative mass from the inflationary field~\cite{ Dine:1995kz}. In this case, $\phi$ is trapped at a much larger value $\phi_0 \gg  H$. Thus, the quantum jump is just  $\delta \theta \sim \frac{H}{2\pi\phi_0}$ which is much smaller than the $\theta_0 \sim \mathcal{O}(1)$. In this case, the correlation length of $\theta$ indeed could be much larger than our current horizon size.  There is also one case that if the Affleck-Dine field plays the role of inflaton, the field value could be much large than the Hubble parameter, then the correlation length could be long and the baryon asymmetry can be generated during inflation~\cite{Barrie:2021mwi, Barrie:2022cub, Han:2022ssz}.

\begin{acknowledgments}
C.H. acknowledges Misao Sasaki, Neil D. Barrie and Shi Pi for their helpful discussions. C. H. is supported by the Guangzhou Basic and Applied Basic Research Foundation under Grant No. 202102020885, and the Sun Yat-Sen University Science Foundation. 
\end{acknowledgments}

\bibliographystyle{JHEP}
\bibliography{new}

\appendix
\section{Numerical calculation of $R_{n,l}$}
Although {\bf Mathematica~12} provides a function ``NDEigensystem" to solve the eigensystem, we can directly apply it to the $R_{n,l}$ function. The reason is that the normalization of the $R_{n,l}$ function is
\begin{eqnarray}
 \int  R_{n,l}(r) R_{n,l}(r) r dr =1~.
\end{eqnarray}
To preform the numerical calculation, we can define  $R_{n,l}=\chi_{n,r} r^{-1/2}$, then the normalization of $\chi$ becomes
\begin{eqnarray}
 \int  \chi_{n,l}(r) \chi_{n,l}(r) dr =1 ~,
\end{eqnarray}
then
\begin{eqnarray}
R^\prime &=& \chi^\prime r^{-1/2}-\frac{1}{2} \chi r^{-3/2}~, \\
R^{\prime \prime} &=& \chi^{\prime \prime} r^{-1/2}- \chi^\prime r^{-3/2}+ \frac{3}{4} \chi r^{-5/2} ~.
\end{eqnarray}
Notice we have
\begin{eqnarray}
- \left[ R^{\prime \prime}_{n,l} + \frac{R^{\prime}_{n,l}}{r} - \frac{l^2}{r^2} R_{n,l} \right]+ 
\left[  (v^\prime)^2 - v^{\prime \prime}-\frac{v^\prime}{r} \right] R_{n,l}=\frac{8\pi^2\Lambda_{n,l}}{H^3} R_{n,l}~,
\end{eqnarray}
and
\begin{eqnarray}
- \left[ \chi^{\prime \prime}_{n,l} + \frac{1}{4r^2} \chi_{n,l}- \frac{l^2}{r^2} \chi_{n,l} \right]+ 
\left[  (v^\prime)^2 - v^{\prime \prime}-\frac{v^\prime}{r} \right] \chi_{n,l}=\frac{8\pi^2\Lambda_{n,l}}{H^3} \chi_{n,l}~.
\end{eqnarray}
Then we can use  $\chi_{n,l}$ to perform all the numerical calculations instead of $R_{n,l}$.
\end{document}